\documentclass[transmag]{IEEEtran}
\usepackage{latexsym}
\usepackage{graphicx}
\usepackage{subfigure}
\usepackage{multirow}
\usepackage{amsfonts,amssymb,amsmath}
\usepackage{hyperref}
\usepackage{bm}
\usepackage{booktabs}
\def\BibTeX{{\rm B\kern-.05em{\sc i\kern-.025em b}\kern-.08em T\kern-.1667em\lower.7ex\hbox{E}\kern-.125emX}}

\usepackage{algorithm} 		
\usepackage{algorithmic} 	
\usepackage{bm}
\usepackage{graphicx}
\usepackage{amsmath} %
\usepackage{flushend}
\usepackage{cite}
\usepackage{amssymb}
\usepackage{multirow}

\usepackage{epstopdf}
\usepackage{exscale}
\usepackage{relsize}
\usepackage{color}
\usepackage{diagbox}
\usepackage{slashbox}
\usepackage{booktabs}
\usepackage{array}
\usepackage{makecell}
\usepackage{multicol}

\newcolumntype{I}{!{\vrule width 1.5pt}}

\newlength\savewidth

\usepackage{subeqnarray}
\usepackage{enumerate}
\usepackage{flushend}
\usepackage{subfigure}

\begin{document}

\title{Two-Layer Stacked Intelligent Metasurfaces: \\Balancing Performance and Complexity}

\author{Hong Niu,~\IEEEmembership{}
        Chau Yuen,~\IEEEmembership{Fellow, IEEE,}
        Marco Di Renzo,~\IEEEmembership{Fellow, IEEE,} \\
        M\'{e}rouane Debbah,~\IEEEmembership{Fellow, IEEE,}
        and H. Vincent Poor,~\IEEEmembership{Life Fellow, IEEE}

\thanks{

Hong Niu and Chau Yuen \emph{(Corresponding author)} are with the School of Electrical and Electronics Engineering, Nanyang Technological University, Singapore 639798 (E-mail: \{hong.niu,chau.yuen\}@ntu.edu.sg).

Marco Di Renzo is with CNRS and CentraleSup\'elec, Institute of Electronics and Digital Technologies, Avenue de la Boulaie, 35576 Cesson-S\'evign\'e, France (E-mail: marco.direnzo@centralesupelec.fr), and with King's College London, Department of Engineering - Centre for Telecommunications Research, WC2R 2LS London, United Kingdom (E-mail: marco.di\_renzo@kcl.ac.uk).


M\'{e}rouane Debbah is with the Research Institute for Digital Future, Khalifa University, 127788 Abu Dhabi, United Arab Emirates (Email: merouane.debbah@ku.ac.ae).

H. Vincent Poor is with the Department of Electrical and Computer Engineering, Princeton University, Princeton, NJ, USA 08544 (Email: poor@princeton.edu).
}
}

\markboth{}
{Shell \MakeLowercase{\textit{et al.}}: }
\maketitle

\begin{abstract}
Stacked intelligent metasurfaces (SIMs) have emerged as a powerful paradigm for wave-domain signal processing, enabling fine-grained control over electromagnetic (EM) propagation in next-generation wireless systems. However, conventional multi-layer SIMs often suffer from excessive structural complexity, high computational overhead, and significant power attenuation across layers, limiting their performance. In this paper, we first characterize SIMs from the perspectives of functionality, application, and layer configuration, revealing the inherent trade-offs between signal processing flexibility and power efficiency. Then, two representative 2-layer architectures, the meta-fiber-connected SIM (MF-SIM) and the flexible intelligent layered metasurface (FILM), are introduced, each advocating a distinct 2-layer SIM design philosophy. Moreover, we identify several open challenges in topology optimization for MF-SIM, shape control for FILM, and hybrid 2-layer architectures. Finally, case studies considering 2-layer MF-SIM and FILM assisted point-to-point multiple-input multiple-output (MIMO) and multi-user communication systems validate that properly designed 2-layer SIMs can significantly reduce power loss and optimization burden while maintaining good signal processing performance, offering a promising pathway toward practical SIM-enabled 6G systems.
\end{abstract}


\IEEEpeerreviewmaketitle

\vspace*{-3mm}
\section{Introduction}

\IEEEPARstart{T}{he} evolution toward sixth-generation (6G) wireless networks is marked by unprecedented demands for ultra-high data rates, minimal latency, massive connectivity, and enhanced spectral and power efficiency. These stringent requirements challenge the capabilities of conventional wireless architectures, prompting a shift toward novel physical-layer solutions \cite{MM1}. Among the most promising options, metamaterial-based technologies enable intelligent control over electromagnetic (EM) waves \cite{Lin1,Qian1}.

Metamaterials, comprising engineered subwavelength elements, offer unique capabilities in manipulating wave propagation, including amplitude, phase, direction, and polarization \cite{Liu1,Gao1}. Leveraging these properties, emerging platforms are redefining the way wireless environments are designed and optimized. Among various metamaterial-based technologies, stacked intelligent metasurfaces (SIMs) offer distinct advantages from a signal processing perspective \cite{SIM1,SIM2}. By layering metasurfaces with tailored transmissive properties, SIMs can manipulate both the amplitude and phase of waves across multiple spatial dimensions, enabling high-resolution waveform synthesis and advanced spatial filtering \cite{SIM3,SIM4}. The multi-layered architecture turns a SIM into a functional signal processor, capable of tasks such as beamforming \cite{SIM5}, channel equalization \cite{SIM1}, and integrated sensing and communication (ISAC) \cite{ISAC1}.

Fundamentally, a SIM employs multiple stacked layers to provide increased degrees of control for enhanced signal processing. However, the optimal design of SIM remains unresolved. Current research trends focus on increasing the number of stacked layers to realize richer theoretical enhancement \cite{SIM1}-\cite{ISAC1}. Nevertheless, this approach raises two practical challenges that may impede SIM deployment. On the one hand, a larger number of layers leads to a higher count of meta-atoms, which significantly increases the number of optimization variables and complicates system optimization. {On the other hand, in practical implementations, signal propagation through multiple layers incurs inevitable power attenuation, where the transmission power decreases with the number of layers, resulting in a rapid deterioration of power efficiency.}

Consequently, for practical SIM implementations, it is critical to balance the signal processing capabilities with the number of required layers \cite{MFR1}-\cite{FILM1}. {Since 1-layer SIMs are generally limited to phase-only control, a 2-layer SIM represents the minimal architecture required to realize independent joint amplitude and phase manipulation in the wave domain.} In this paper, we aim to shed light on a promising direction for future 2-layer SIM architecture design. This paradigm not only reduces the task of SIM optimization but also paves the way for energy-efficient SIM implementations to meet the demands of 6G systems. This work represents a significant step toward realizing 2-layer SIMs and highlights the potential for innovation in balancing SIM performance and complexity.

\begin{figure*}[!htb]
\centering
\includegraphics[width=\linewidth]{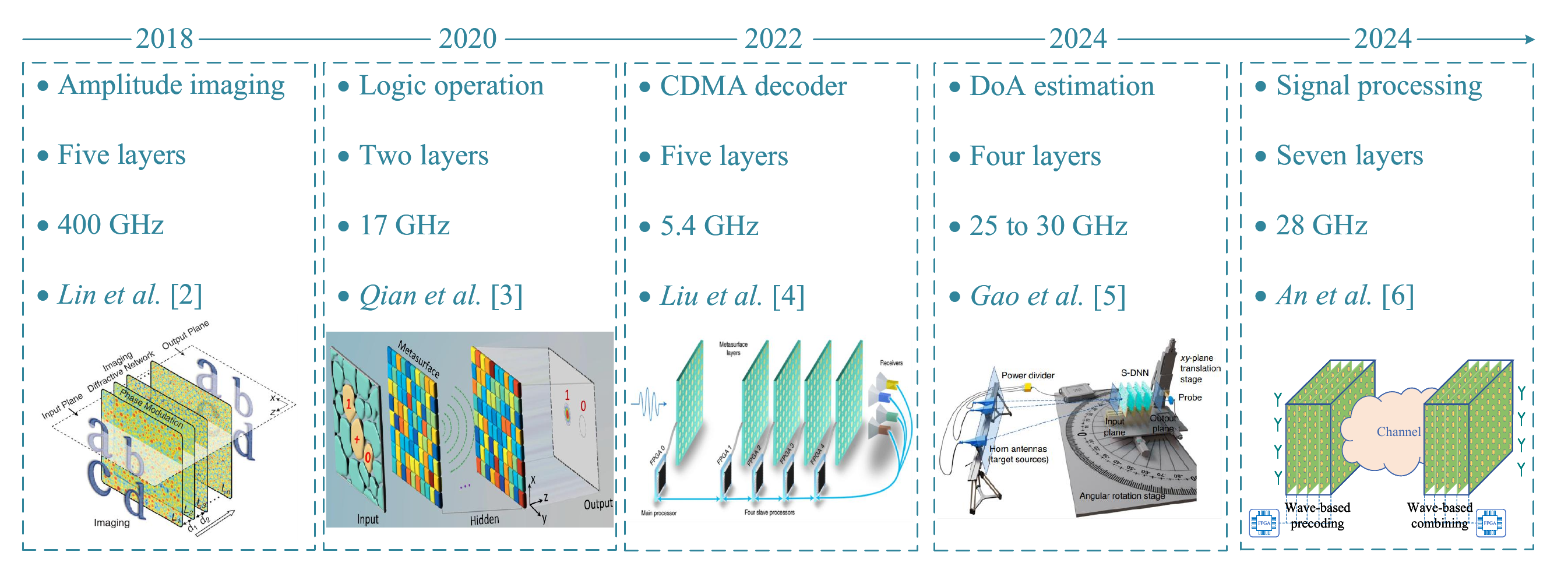}
\caption{{Representative classification and functional overview of SIM architectures.}}
\label{fig_1}
\vspace{+1em}
\end{figure*}

\section{Characterization of SIM}

Unlike conventional reconfigurable intelligent surfaces (RISs) that primarily manipulate reflected EM waves, SIMs operate in transmissive mode, extending RIS functionality by introducing multiple reconfigurable layers for enhanced wave control. By stacking multiple reconfigurable metasurface layers, SIMs offer enhanced control over amplitude, phase, polarization, and spatial dispersion, acting as three-dimensional signal processors.

Several prototypes of SIMs, as illustrated in Fig. 1, have demonstrated their feasibility using varied designs. This section characterizes SIMs from the perspectives of functionality, bandwidth, and the number of layers, revealing key trade-offs and trends to guide future wireless system developments.

\subsection{Functional Capabilities}

{The strategic design of SIMs is guided by their intended functionalities within wireless communication and sensing ecosystems. These functionalities can be broadly classified into three main categories:}

\textbf{F1: Communication Capabilities.} SIMs enable advanced wave-domain communications by jointly controlling amplitude and phase across layers. They can realize beamforming, spatial filtering, modulation, and multi-user separation while dynamically reconfiguring the propagation environment. Acting as programmable processors, SIMs support adaptive, spectrum-efficient, and secure transmission for multiple-input multiple-output (MIMO) systems.

\textbf{F2: Sensing Capabilities.} By tuning their transmissive properties, SIMs can sense fine-grained environmental information such as direction, distance, and spatial distribution. This allows direct realization of angle estimation and localization without extra sensors.

\textbf{F3: Computation Capabilities.} Inter-layer EM interactions give SIMs intrinsic analog computing capabilities. Properly designed transmission coefficients enable complex operations, such as the computation of the two-dimensional discrete Fourier transform (2D DFT). Hence, SIMs can be viewed as energy-efficient in-hardware (or in-metasurface) computing platforms.

\begin{table*}[!htb]
\large
\center
\caption{{Representative research works on SIM-enabled wave-domain processing.}}\label{tabI}
\vspace{+5mm}
\renewcommand\arraystretch{1}
\resizebox*{\linewidth}{!}{
\begin{tabular}{|>{\centering\arraybackslash}m{1cm}|>{\centering\arraybackslash}m{2cm}|>{\centering\arraybackslash}m{2cm}|>{\centering\arraybackslash}m{2cm}|m{10.5cm}|}
\hline
\textbf{Refs.} & {\textbf{Function}} & {\textbf{Application}} & \textbf{Number} & \textbf{Key contributions} \\ \hline
\cite{MM1} & F1, F2 & A1-A5 & N1-N3 & Overview SIM as a novel technology enabling fast, low-latency EM signal processing for next-generation wireless communications. \\ \hline
\cite{Lin1} & F2 & A5 & N1 & Introduce an all-optical SIM that performs image classification through passive diffractive layers, enabling light-speed optical computation. \\ \hline
\cite{Qian1} & F3 & A5 & N3 & Propose a diffractive neural network-based SIM that performs basic logic operations using simple plane-wave inputs, eliminating the need for complex optical control.\\ \hline
\cite{Liu1} & F2 & A5 & N2 & Present an active SIM, enabling reconfigurable all-optical intelligence for CDMA decoding.\\ \hline
\cite{SIM1} & F1 & A2 & N1 & Jointly design the phase shifts of SIMs at the transmitter and receiver to implement parallel subchannels without excessive RF chains. \\ \hline
\cite{SIM2} & F1, F2 & A1-A4 & N1 & Overview SIM-aided MIMO transceiver design, hardware architecture, potential benefits, and open challenges. \\ \hline
\cite{SIM3} & F1 & A2 & N1 & Introduce a SIM-assisted cell-free massive MIMO framework to reduce system complexity and enhance sum-rate. \\ \hline
\cite{SIM4} & F2, F3 & A3 & N1 & Design SIM to automatically perform the 2D DFT and estimate 2D direction-of-arrival of incident signals. \\ \hline
\cite{SIM5} & F1 & A2 & N1 & Propose a fully analog SIM architecture for MIMO-OFDM systems that performs beamforming to eliminate inter-antenna interference and enhance channel capacity. \\ \hline
\cite{ISAC1} & F1, F2 & A4 & N1, N2 & {Jointly optimize BS beamforming and SIM configuration for Cram\'{e}r-Rao bound minimization in a SIM-assisted ISAC system, validated via hardware experiments.} \\ \hline
\cite{MFR1} & F1 & A2, A5 & N2 & Demonstrate a nonlinear active electronic-photonic SIM that converts optical signals into steerable millimeter-wave beams, enabling high-speed fiber-wireless communication. \\ \hline
\cite{SIM6} & F1 & A2 & N3 & Introduce meta-fiber to achieve a low-complexity design with only two layers. \\ \hline
\cite{FIM2} & F1, F2 & A1-A4 & N1-N3 & Overview FIM and SIM technologies, discussing its impact on wireless communications and identifying key challenges for future research. \\ \hline
\cite{FILM1} & F1 & A2 & N3 & Introduce a 2-layer FILM architecture that dynamically adjusts transmission coefficients. \\ \hline
\end{tabular}
}
\end{table*}

\subsection{Applications}

The applications of SIM-assisted systems are diverse and continue to expand with ongoing research.
Representative applications can be broadly categorized into the following five domains:

\textbf{A1: Physical-Layer Security.} SIMs have been leveraged to enhance physical-layer security by generating artificial noise through controllable phase shifts, thereby confusing potential eavesdroppers without relying on upper-layer encryption.

\textbf{A2: MIMO.} Several studies have exploited SIMs to enable high-resolution MIMO transmissions, including point-to-point MIMO, wideband orthogonal frequency-division multiplexing (OFDM), and multi-user communications. These implementations improve spectral efficiency and energy utilization while reducing hardware complexity.

\textbf{A3: Localization.} SIMs have been utilized to facilitate high-accuracy positioning by tailoring the array responses to estimate the direction of arrival (DoA) and spatial location of incident signals, thereby enhancing localization performance in multipath environments.

\textbf{A4: ISAC.} SIMs allow the joint optimization of signal reflection and beam patterns, enhancing target detection and communication quality simultaneously. Such flexible waveform shaping supports the dual-functionality of sensing and communication.

\textbf{A5: Optical Processing.} {Optical SIMs have been successfully implemented and applied for image classification, logic computation, and code division multiple access (CDMA) decoding, enabling light-speed information processing.}

%
%
%

\subsection{Number of Layers}

SIMs can be categorized based on the number of stacked layers into multi-layer, medium-layer, and 2-layer configurations, each presenting unique trade-offs in design complexity, performance, and practical deployment.

\textbf{N1: Multi-Layer Structures.} Multi-layer SIMs, typically with more than five layers, offer abundant degrees of control for advanced wave manipulation, enabling high-resolution beamforming and complex spatial filtering. However, their deep stacking causes severe power attenuation, high optimization complexity, and significant fabrication challenges, making large-scale deployment costly and difficult.

\textbf{N2: Medium-Layer Structures.} Medium-layer SIMs with three to five layers balance performance and practicality. They provide richer signal processing than 1-layer designs while alleviating part of the power loss and computational burden of deeper structures. Nonetheless, power attenuation and design complexity remain notable issues, limiting further scalability.

\textbf{N3: Two-Layer Structures.} 2-layer SIMs emphasize simplicity and efficiency. While 1-layer SIMs or RISs mainly offer phase control, 2-layer SIMs represent the minimal configurations to achieve joint amplitude-phase enabled wave-domain processing. The 2-layer architectures minimize power loss and hardware cost, making them ideal for practical and early-stage 6G applications.

Table \ref{tabI} summarizes recent advances in SIM-assisted wireless communication systems, categorized by functionality, frequency band, and layer count. While multi-layer SIMs provide enhanced beamforming and signal processing capabilities, they suffer from increased power attenuation and significantly greater optimization and fabrication complexity. Therefore, focusing on 2-layer SIMs offers a balanced trade-off among performance, power efficiency, fabrication cost, and ease of integration, making them ideal for practical deployment and early-stage applications.

{Next, we will present design strategies for two representative 2-layer SIM architectures.}

\section{Design Concept of Two-Layer SIM}

\begin{figure}[t]
\centering
\includegraphics[width=\linewidth]{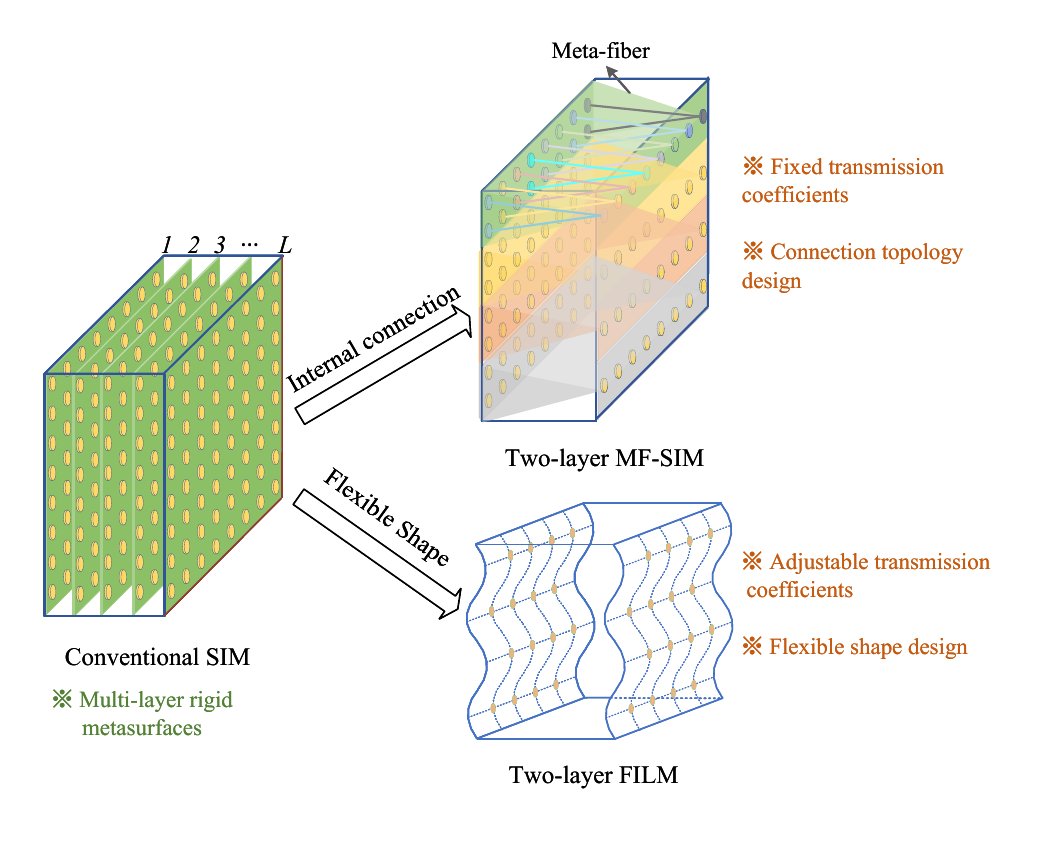}
\caption{Two design strategies for two-layer SIMs.}
\label{fig_2}
\vspace{+1em}
\end{figure}

{The 2-layer meta-fiber-connected SIM (MF-SIM) and the 2-layer flexible intelligent layer metasurface (FILM), as illustrated in Fig. \ref{fig_2}, represent two innovative approaches for designing SIM architectures with enhanced wave-domain signal processing capabilities.}

\subsection{MF-SIM Design Philosophy}

The 2-layer MF-SIM is developed to address the inherent limitations of conventional multi-layer SIMs. {In conventional multi-layer SIMs, the core degrees of control required for wave-domain signal processing, i.e., the independent manipulation of the output signal amplitude and phase, are typically achieved through uniform stacking of multiple layers. This design inevitably results in increased computational complexity and severe power attenuation due to the cascaded wireless propagation. By contrast, the proposed 2-layer MF-SIM constrains the architecture to the minimal number of layers while preserving the same level of flexibility in joint amplitude-phase control.} Moreover, by converting inter-layer signal transmission from wireless propagation to deterministic wired connections via meta-fibers, MF-SIM achieves simplified optimization, reduced power loss, and enhanced implementation feasibility.

{A key innovation of the MF-SIM lies in the introduction of meta-fibers, which act as fixed transmission coefficients between meta-atoms across layers. As demonstrated in \cite{MFR1}, meta-fibers can be implemented using an electronic-photonic metasurface platform that converts an incident optical wave at 193 THz into a millimeter-wave signal at 28 GHz, thus bridging optical and RF domains. Unlike conventional SIMs whose transmission coefficients are modeled by Rayleigh-Sommerfeld propagation \cite{SIM1}, the use of meta-fibers enables direct control, improving signal fidelity and reducing inter-layer interference. This controllable interconnection also provides a flexible internal topology analogous to perceptron layers in neural networks.

In the 2-layer MF-SIM, each layer is partitioned into sub-areas connected via meta-fibers following a 2-to-1 partial topology \cite{SIM6}. The first-layer meta-atoms primarily modulate the amplitudes of signals, while the second layer controls the phases of signals. This coordinated design enables the flexible synthesis of spatial waveforms with arbitrary amplitude and phase distributions, achieving efficient wave-domain signal processing with minimal structural complexity.}

\subsection{FILM Design Philosophy}

In contrast, the 2-layer FILM adopts a fundamentally different strategy for adjusting transmission coefficients, by not requiring inter-layer wiring, but physically reconfiguring the positions of meta-atoms within each layer. Each layer of the FILM consists of a planar array of meta-atoms capable of mechanical displacements compared to a rigid reference configuration. This dynamic morphing changes the transmission distances and angles between meta-atoms on adjacent layers, thereby modulating the transmission coefficients in a controllable manner \cite{FILM1}.

Unlike the fixed topology of the MF-SIM, the 2-layer FILM exploits spatial displacement of meta-atoms to introduce additional degrees of control. The inter-layer transmission follows near-field propagation and diffraction principles, but its behavior can be dynamically altered through shape deformation of the metasurface. {The FILM architecture can be realized through either mechanically driven approaches or electromagnetic actuation, such Lorentz-force-based control \cite{FIM2,FILM1}.}

It is worth mentioning that the channel gain variation induced by the displacements of meta-atoms is relatively limited in 1-layer flexible intelligent metasurface (FIM) designs \cite{FIM2}, as small changes in the positions of the meta-atoms have little effect in large-scale, low-scattering environments. In the 2-layer FILM configuration, on the other hand, even small adjustments of the transmission coefficients can significantly influence signal processing performance, owing to the high sensitivity to phase variations.

\subsection{Comparison of MF-SIM and FILM}

\begin{table*}[!htb]
\large
\center
\caption{Comparison among two-layer MF-SIM, FILM, and conventional multi-layer SIM.}\label{tabII}
\vspace{+5mm}
\renewcommand\arraystretch{1}
\resizebox*{\linewidth}{!}{
\begin{tabular}{|>{\centering\arraybackslash}m{1.5cm}|>{\centering\arraybackslash}m{2cm}|>{\centering\arraybackslash}m{2cm}|>{\centering\arraybackslash}m{2.5cm}|>{\centering\arraybackslash}m{3cm}|>{\centering\arraybackslash}m{1.5cm}|m{6.5cm}|}
\hline
\textbf{Schemes} & {\textbf{Number of Layers}} & {\textbf{Structural Control}} & \textbf{Inter-layer Transmission} & \textbf{Computational Complexity} & \textbf{Power Loss} & \textbf{Distinctive Features} \\ \hline
SIM & Multi-layer & Fixed & Wireless & High & High & Performs wave-domain signal processing through wireless propagation across multiple diffractive layers. \\ \hline
MF-SIM & Two-layer & Fixed & Wired & Low & Low & Employs a meta-fiber-based wired topology that connects two layers. \\ \hline
FILM & Two-layer& Dynamic & Wireless & Moderate & Moderate & Employs a flexible metasurface shape to dynamically adjust transmission coefficients across two layers. \\ \hline
\end{tabular}
}
\end{table*}

{Both the MF-SIM and the FILM represent innovative SIM architectures that enable wave-domain signal processing with reduced computational complexity and power loss.} However, their operating principles differ fundamentally, as summarized in Table \ref{tabII}.

\textbf{Structural Control Mechanism:} The MF-SIM relies on fixed inter-meta-atom connections through meta-fibers, where the transmission coefficient is determined by the internal topological connection. Once fabricated, these transmission coefficients remain fixed, ensuring stable and deterministic inter-layer signal transmission.

In contrast, the FILM adjusts the transmission coefficients by controlling the shape of metasurfaces, granting additional degrees of control for dynamic configuration of inter-layer propagation characteristics.

\textbf{Computational Complexity:} Due to its fixed inter-layer connection and limited number of meta-atoms, the MF-SIM exhibits low computational complexity in system optimization and signal reconstruction. The FILM introduces moderate complexity, as its continuously adjustable surface shape requires iterative calibration of transmission coefficients. In comparison, conventional multi-layer SIMs suffer from high complexity due to the large number of meta-atoms and multiple diffractive layers.

\textbf{Power Loss:} In an MF-SIM, guided wired transmission between two layers minimizes inter-layer attenuation, resulting in low power loss and high signal fidelity. A FILM, by contrast, operates via near-field wireless propagation between two layers, leading to moderate power loss. Conventional multi-layer SIMs experience the highest power attenuation during multi-stage wireless propagation.

Overall, the MF-SIM provides a straightforward approach to simplify SIM design through deterministic wired interconnections, while the FILM emphasizes flexibility through reconfigurable metasurface shapes. These two architectures offer complementary design paradigms for future SIM development, enabling a flexible balance between performance, complexity, and power efficiency.

\section{Research Challenges for Two-Layer SIMs}

The emergence of 2-layer MF-SIMs and FILMs introduces new opportunities for wave-domain signal processing with reduced complexity and power loss, while also raising unique research challenges that differ fundamentally from conventional multi-layer SIM designs. Here, we highlight three key research directions and associated challenges, along with potential approaches to address them.

\subsection{Topology Optimization for MF-SIMs}

Unlike multi-layer SIMs, where inter-layer transmission is governed by wireless diffraction, MF-SIMs rely on fixed meta-fiber connections. Designing optimal topological mappings between meta-atoms across two layers to achieve target amplitude-phase waveforms while minimizing hardware usage is nontrivial. The use of fixed wires imposes constraints on flexibility, making conventional gradient-based optimization insufficient.

{Graph-based topology optimization or combinatorial algorithms may be employed to construct meta-fiber connections that satisfy the design objective of MF-SIM, namely enabling arbitrary output amplitude and phase synthesis under a given topology. This objective focuses on the reachability of desired waveforms rather than the minimization of a single scalar cost function, as demonstrated in \cite{SIM6}. In this context, machine learning techniques, such as reinforcement learning, may serve as heuristic tools to explore large topology design spaces and identify effective signal routing and amplitude-phase control strategies.}

\subsection{Shape Control and Sensitivity Management for FILMs}

FILMs achieve dynamic control through mechanical displacement of meta-atoms, introducing additional degrees of control for signal processing. However, precise control of small-scale deformations is challenging, as minor positioning errors can significantly impact transmission coefficients and phase accuracy. This sensitivity is largely absent in conventional multi-layer SIMs, where the metasurface layers are rigid. {Moreover, displacement inaccuracies can be modeled as spatial perturbations that translate into errors in transmission coefficients, leading to performance degradation.}

{One possible approach to mitigate positioning errors in FILMs is to employ closed-loop feedback control, where sensing or monitoring mechanisms are used to adjust meta-atom positions during operation. The achievable real-time response and phase correction accuracy of such feedback systems depend on implementation-specific factors and therefore remain open questions. In addition, optimization and learning-based calibration strategies may be explored to improve robustness against mechanical tolerances and positioning errors.}

\subsection{Hybrid Two-Layer Architectures}

While multi-layer SIMs rely on stacking several metasurface layers to increase the degrees of control, 2-layer designs must achieve comparable functionalities with only two layers. {Balancing the trade-off between signal processing capability and power efficiency is critical, as a wired MF-SIM reduces attenuation but may limit flexibility under simplified topologies, whereas a FILM offers flexibility at the cost of additional power losses.

A hybrid 2-layer architecture may capitalize on these complementary features by employing meta-fiber-assisted wired connections for inter-layer coupling, while enabling meta-atom position reconfiguration at the input and output layers. In this manner, the hybrid design introduces additional controllable variables to enhance flexibility beyond MF-SIM, while mitigating the power loss inherent in FILM by removing the inter-layer wireless channel. Although such an architecture poses nontrivial implementation challenges, it represents a promising research direction for balancing flexibility and power efficiency.

In summary, these considerations highlight the distinct yet complementary characteristics of 2-layer SIMs: MF-SIM's fixed wired topology, FILM's mechanically adjustable shape, with the hybrid architectures combining both. The development of detailed coordination mechanisms among system components, together with a quantitative analysis of the complexity-performance trade-off, remain important directions for future implementation-oriented studies of hybrid 2-layer architectures.}

\section{Case Study}

In this section, we present two case studies to demonstrate the benefits of 2-layer MF-SIMs and FILMs in point-to-point MIMO and multi-user communications.

\subsection{Point-to-Point MIMO Communications}

{We consider a SIM-assisted point-to-point MIMO system, where the transmitter employs a SIM for wave-domain precoding}. Each meta-atom within the SIM is capable of phase modulation only, without amplitude control. The free-space propagation channel between the transceiver pair is modeled as a complex Gaussian Rayleigh channel.

To evaluate system performance, simulations are conducted for the transmission of four data streams at a 28 GHz carrier frequency, with a transmit power of 20 dBm and additive white Gaussian noise of $-110$ dBm. The MIMO channel assumes a path loss exponent of 2.5 over a 150 m link distance. Each layer of the proposed 2-layer SIM contains 100 meta-atoms. For simplicity, the meta-fibers are assumed ideal, introducing no amplitude loss. {To partially account for practical hardware impairments, a per-layer power attenuation factor introduced by the metasurfaces is considered.} For a fair comparison, the following benchmark schemes are considered:

{
1) \textbf{2-layer MF-SIM}: A 2-layer MF-SIM structure that employs a 2-to-1 partial topology connection, where the inter-layer transmission coefficients are deterministically defined by fixed connections, as considered in \cite{SIM6}.

2) \textbf{2-layer FILM}: A morphable 2-layer architecture in which the wireless inter-layer transmission coefficients are modeled according to Rayleigh-Sommerfeld diffraction propagation theory and are controllable via physical displacement of meta-atoms, as considered in \cite{FILM1}.

3) \textbf{Conventional multi-layer SIM}: A conventional uniformly arranged multi-layer structure with wireless inter-layer coupling modeled according to Rayleigh-Sommerfeld diffraction propagation theory, in which only the phase shifts are adjustable. This category is distinct from MF-SIM and FILM. As case studies, we consider 7-layer, 4-layer, and 1-layer configurations, as in \cite{SIM1}.}

4) \textbf{MIMO}: A conventional digital beamforming system with the same number of RF chains but without metasurface assistance.

\begin{figure}[t]
\centering
\includegraphics[width=\linewidth]{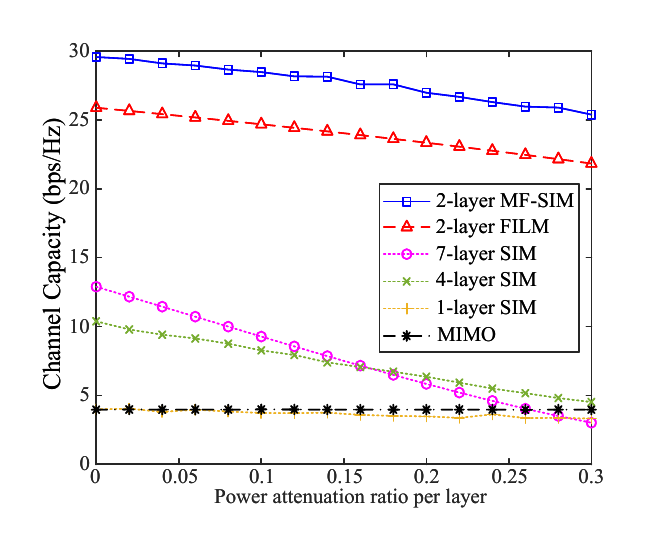}
\caption{Channel capacity versus the power attenuation ratio per layer in point-to-point MIMO systems.}
\label{fig_4}
\end{figure}

Our analysis focuses on the power attenuation ratio per layer, i.e., the signal power loss experienced through each metasurface layer, as illustrated in Fig. \ref{fig_4}. As expected, the channel capacity decreases when the power attenuation increases. The conventional MIMO system remains unaffected since no metasurface is utilized. {Among the metasurface-based schemes, the 7-layer SIM exhibits the most pronounced performance degradation. The degradation rate is followed by the 4-layer, the 2-layer, and finally the 1-layer SIM. This behavior arises from the fact that the output power efficiency decreases with the number of layers, where each additional layer introduces a fixed power attenuation factor. When the power attenuation ratio exceeds 17\%, 26\%, and 28\%, the performance of a 7-layer SIM drops below that of a 4-layer SIM, conventional MIMO, and 1-layer SIM, respectively.}

In contrast, both the 2-layer MF-SIM and 2-layer FILM demonstrate significantly improved resilience to the inter-layer power loss. The MF-SIM benefits from its fixed and efficient topology, reinforced by the use of meta-fiber interconnections, which effectively reduce the transmission loss. The FILM architecture, on the other hand, leverages its morphable geometry to optimize propagation paths and maintain superior power utilization.

\subsection{Multi-User Communications}

\begin{figure}[t]
\centering
\includegraphics[width=\linewidth]{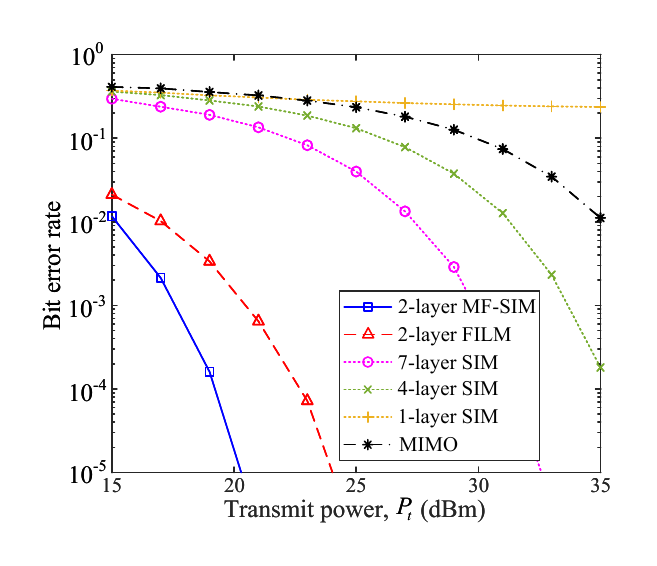}
\caption{Bit-error rate versus the transmit power in multi-user MISO systems.}
\label{fig_5}
\end{figure}

{We consider a SIM assisted multi-user multiple-input single-output (MISO) system, where the transmitter employs a SIM to perform wave-domain beamforming.} Each meta-atom within the SIM can modulate only the phase of the incident EM wave. {The communication channel between the 2-layer SIM and the users is characterized by transmit-side steering vectors, which capture the spatial responses of the flexible metasurface toward each user direction, incorporating both the phase and displacement effects of the meta-atoms.}

Simulations are conducted for four users at a 28 GHz carrier frequency, assuming 30 dBm transmit power and $-125$ dBm additive white Gaussian noise per receiver. The path-loss exponent is 2.5, and users are located 150 m far from the FILM at distinct elevation and azimuth angles. {The BS employs as many RF chains as the number of users, uses quadrature phase shift keying modulation, and the antenna array (half-wavelength spacing) is centered and aligned with the FILM.} The 2-layer FILM comprises two $10\times10$ meta-atom layers (half-wavelength spacing), separated by 5 mm with a maximum morphing range of 2.4 mm. Inter-layer propagation follows the Rayleigh-Sommerfeld model.

Simulation results show that, for a target bit-error rate of $10^-5$, the 2-layer FILM and MF-SIM architectures reduce the required transmit power by over 7 dB and 11 dB, respectively, compared to a 7-layer SIM. The improvement arises from FILM's dual control of phase and shape, enabling adaptive propagation shaping and effective multi-user interference suppression. {As expected, in conventional multi-layer SIM schemes, the 7-layer SIM outperforms the 4-layer SIM, which in turn outperforms the 1-layer SIM. Due to its inability to achieve ideal signal processing, the 1-layer SIM exhibits performance inferior to that of the MIMO system assisted by digital beamforming.}

In summary, both 2-layer MF-SIM and FILM strike an appealing balance between architectural simplicity and high performance.

\section{Conclusion}

This paper has put forth 2-layer SIM as a pragmatic technology to balance performance and complexity in 6G wireless systems. By revisiting the SIM design philosophy from the perspective of structural simplicity and power efficiency, we have highlighted that reducing the number of layers does not necessarily compromise functionality but instead enhances system deployability and energy sustainability. Through the examination of two representative architectures, the 2-layer MF-SIM and FILM, we have demonstrated that 2-layer SIMs can retain powerful wave-domain processing capabilities while substantially alleviating optimization complexity and power attenuation. {In practical deployments, multi-cell interference, mobility, and frequency selectivity mainly affect channel dynamics and configuration robustness, and their effects similarly impact both 2-layer and conventional multi-layer SIMs.} Future research should focus on topology optimization for MF-SIM, shape control and sensitivity for FILM, and hybrid 2-layer architectures to unlock the full potential of 2-layer metasurface-assisted networks. Overall, this study lays theoretical and architectural groundwork for the next generation of energy-efficient and scalable SIM-based wireless systems.

\section*{Acknowledgments}

This work was supported in part by the  Ministry of Education (MOE), Singapore, under its MOE Tier 2 (Award number T2EP50124-0032). The work of M. Di Renzo was supported in part by the European Union through the Horizon Europe project COVER under grant agreement number 101086228, the Horizon Europe project UNITE under grant agreement number 101129618, the Horizon Europe project INSTINCT under grant agreement number 101139161, and the Horizon Europe project TWIN6G under grant agreement number 101182794, as well as by the Agence Nationale de la Recherche (ANR) through the France 2030 project ANR-PEPR Networks of the Future under grant agreement NF-YACARI 22-PEFT-0005, and by the CHIST-ERA project PASSIONATE under grant agreements CHIST-ERA-22-WAI-04 and ANR-23-CHR4-0003-01. Also, the work of M. Di Renzo was supported in part by the Engineering and Physical Sciences Research Council (EPSRC), part of UK Research and Innovation, and the UK Department of Science, Innovation and Technology through the CHEDDAR Telecom Hub under grant EP/X040518/1 and grant EP/Y037421/1, and through the HASC Telecom Hub under grant EP/X040569/1.

\ifCLASSOPTIONcaptionsoff
  \newpage
\fi

\textsc{Hong Niu} received the B.E. and Ph.D. degrees from the University of Electronic Science and Technology of China, Chengdu, China, in 2018 and 2024, respectively. He is currently a Research Fellow with the School of Electrical and Electronic Engineering, Nanyang Technological University, Singapore. His research interests include physical-layer security, wireless communications, and quantum communications.

\textsc{Chau Yuen} (Fellow, IEEE) received the B.Eng. and Ph.D. degrees from Nanyang Technological University, Singapore, in 2000 and 2004, respectively. He is currently a Provost's Chair in Wireless Communications with the School of Electrical and Electronic Engineering, Nanyang Technological University, Singapore, where he also serves as an Assistant Dean in the Graduate College and Cluster Director for Sustainable Built Environment at ER@IN. His research interests include wireless communications, network intelligence, and sustainable systems. He is a Highly Cited Researcher by Clarivate and listed among the Top 2\% Scientists by Stanford University. He received several awards, including the IEEE Communications Society Leonard G. Abraham Prize and the IEEE Marconi Prize Paper Award in Wireless Communications.

\textsc{Marco Di Renzo} (Fellow, IEEE) is Chair Professor of Telecommunications Engineering, the Director of the Centre for Telecommunications Research, and the Head of the Telecommunications Group, Department of Engineering, King's College London, London, United Kingdom. He is also a CNRS Research Director with the Institute of Electronics and Digital Technologies at CentraleSup\'elec, Rennes, France. He has received several distinctions, including the Michel Monpetit Prize conferred by the French Academy of Sciences, the IEEE Communications Society Heinrich Hertz Award, and the IEEE Communications Society Marconi Prize Paper Award in Wireless Communications. He served as the Editor-in-Chief of IEEE Communications Letters from 2019 to 2023, and as the Director of Journals and Chair of the Publications Misconduct Ad Hoc Committee of the IEEE Communications Society from 2024 to 2025.

\textsc{M\'{e}rouane Debbah} (Fellow, IEEE) is a Professor at Khalifa University of Science and Technology in Abu Dhabi and Founding Senior Director of the KU Digital Future Institute. He is an IEEE Fellow, a WWRF Fellow, a EURASIP Fellow, an AAIA Fellow, an Institut Louis Bachelier Fellow, an AIIA Fellow, and a Membre \'em\'erite SEE. He is currently Chair of the IEEE Large Generative AI Models in Telecom (GenAINet) Emerging Technology Initiative and a member of the Marconi Prize Selection Advisory Committee.

\textsc{H. Vincent Poor} (Life Fellow, IEEE) is the Michael Henry Strater University Professor at Princeton University, where his interests include information theory, stochastic analysis and machine learning, and their applications in wireless networks, energy systems, and related fields. Among his publications in these areas is the book Machine Learning and Wireless Communications (Cambridge University Press, 2022). Dr. Poor is a member of the U.S. National Academy of Engineering and the U.S. National Academy of Sciences, and he a foreign member of the Royal Society and other national and international academies. He received the IEEE Communication Society's Edwin Howard Armstrong Achievement Award in 2009, and the IEEE Alexander Graham Bell Medal in 2017.

\end{document}